\documentclass[twocolumn,superscriptaddress,floatfix,showpacs,aps]{revtex4}%
\usepackage[english]{babel}
\usepackage[utf8]{inputenc}
\usepackage{amsmath}
\usepackage{graphicx,epstopdf,enumerate}
\usepackage[colorinlistoftodos]{todonotes}
\usepackage{amssymb,epsfig,color,textcase}
\usepackage{hyperref}
\usepackage{doi}
\usepackage{amsfonts}
\usepackage{amssymb}
\usepackage{graphicx}%
\providecommand{\U}[1]{\protect\rule{.1in}{.1in}}
\providecommand{\U}[1]{\protect\rule{.1in}{.1in}}

\begin{document}
\title{Spin-orbit driven Peierls transition and possible exotic superconductivity in
CsW$_{2}$O$_{6}$}
\author{Sergey V.~Streltsov}
\email{streltsov@imp.uran.ru}
\affiliation{M.N. Miheev Institute of Metal Physics of Ural Branch of Russian Academy of
Sciences, 620137, Ekaterinburg, Russia}
\affiliation{Ural Federal University, Mira St. 19, 620002 Ekaterinburg, Russia}
\author{Igor I. Mazin}
\affiliation{Code 6393, Naval Research Laboratory, Washington, DC 20375, USA}
\author{Rolf Heid}
\affiliation{Forschungszentrum Karlsruhe, Institut f\"ur Festk\"orperphysik, P.O.B. 3640, D-76021 Karlsruhe, Germany}
\author{Klaus-Peter Bohnen}
\affiliation{Forschungszentrum Karlsruhe, Institut f\"ur Festk\"orperphysik, P.O.B. 3640, D-76021 Karlsruhe, Germany}
\date{\today}

\begin{abstract}
We study \textit{ab initio} a pyrochlore compound, CsW$_{2}$O$_{6}$, which
exhibits a yet unexplained metal-insulator transition. We find that (1) the
reported low-$T$ structure is likely inaccurate and the correct structure has
a twice larger cell; (2) the insulating phase is not of a Mott or
dimer-singlet nature, but a rare example of a 3D Peierls transition, with a
simultaneous condensation of three density waves; (3) spin-orbit interaction
plays a crucial role, forming well-nested bands. The high-$T$ (HT) phase, if
stabilized, could harbor a unique $e_{g}+ie_{g}$ superconducting state that
breaks the time reversal symmetry, but is not chiral. This state was predicted
in 1999, but never observed. We speculate about possible ways to stabilize the
HT phase while keeping the conditions for superconductivity.

\end{abstract}

\maketitle

\textit{Introduction}. Insulating transition metal compounds with partially
filled $d$ shell most often turn out to be locally magnetic (either forming a
long-range magnetic order or remaining paramagnetic). There are some
exceptions to this rule when, for instance, a transition metal ion is in the
low-spin configuration due to a strong crystal-field
splitting\cite{khomskii2014transition}. Alternatively, a Haldane state may
appear in low-dimensional materials with integer spins, as e.g. in Tl$_{2}%
$Ru$_{2}$O$_{7}$\cite{Lee2006}. Yet another possibility is formation of
spin-singlet molecular clusters, such as dimers and timers (as it happens in
VO$_{2}$\cite{khomskii2014transition}, Li$_{2}$RuO$_{3}$
\cite{Miura2007,Kimber2013,Streltsov2014,Wang2014} or Ba$_{4}$Ru$_{3}$O$_{10}$
\cite{Klein2011,Streltsov2012a}) or even more complex objects (heptamers in
AlV$_{2}$O$_{4}$\cite{Horibe2006} or octamers in CuIr$_{2}$S$_{4}%
$\cite{Radaelli2002}). However, in order to get a singlet ($S=0$) ground state
one needs an even number of electrons (doubly) occupying lowest energy levels
in such clusters, as it occurs in Li$_{2}$RuO$_{3}$ or AlV$_{2}$O$_{4}$. In
this sense a recent discovery of zero magnetic susceptibility in the
insulating $\beta-$pyrochlore CsW$_{2}$O$_{6},$ with average occupancy 1/2
electron per site, looks very unusual\cite{Hirai2013}.

This compound undergoes a metal-insulator transition at 210 K with a
Pauli-like magnetic susceptibility for $T>210$ K, while at lower temperatures,
in an insulating phase, it is fully nonmagnetic \cite{Hirai2013}. The high
temperature (HT) phase is cubic (space group $Fd3m$)\cite{Cava1993}. A
complicated structure was proposed for the low temperature (LT) phase, with a
doubled unit cell (compared to the FCC Bravais lattice of two formula units),
with a disproportionation into two types of W and three types of W-W bonds.
The short bonds form 1D zigzag chains\cite{Hirai2013}. At the same time, the
average W-O distance (the valence bond sum) is nearly the same for both W,
indicating absence of a charge order. Obviously, uniform 1D chains with
noninteger number of electrons per site cannot form a simple band insulator.

The insulating and nonmagnetic nature of the low-temperature phase of
CsW$_{2}$O$_{6}$, given the absence of charge disproportionation, no di- or
tetramer formation, and 1/2 electron per metal site, remains mysterious. Most
usual suspects for explaining such transitions patently fail in CsW$_{2}
$O$_{6}.$ Indeed,

(i) Strong spin-orbit coupling (SOC), typical for $5d$ metals such as W, in
principle may stabilize a nonmagnetic state with the orbital moment
antiparallel to spin and the total moment $J=0.$ However, while this may be
the case for $d^{4}$ configuration\cite{Abragam,khomskii2014transition}, it is
not possible for the $d^{1/2}$ occupancy. Besides, this model cannot explain
the insulating behavior.

(ii) Correlation effects such as a Mott-Hubbard transition with possible
formation of spin-singlets below 210 K would result in formation of local spin
moments, manifestly absent at any temperature.

(iii) In principle, exotic electron-phonon coupling could stabilize
bipolarons, whereupon every fourth W would have a nonmagnetic $d^{2}$
configurations, and all others the nonmagnetic $d^{0}$. However, that would
generate a considerable O breathing distortion around the $d^{2}$ atom, which
would be hardly possible to miss in the experiment. Besides, that would have
to work against the Hubbard $U,$ which, while small in W, would still amount
to at least 1 eV.

We are left with the only possible scenario: the metal-insulator transition
here is of the Peierls type, and the low-T state is a band insulator. The
reported low-T structure \cite{Hirai2013} does show symmetry lowering: along
the W-W nearest neighbor directions the bonds alternate as
short-short-long-long, reminiscent of the proposition of Mizokawa and
Khomskii\cite{Khomskii2005a}, who suggested that, in analogy with MgTi$_{2}%
$O$_{4},$ in CuIr$_{2}$S$_{4}$ 1/2 hole per metal can form a quasi-1D band
along the Ir-Ir bonds, resulting in a Peierls transition with tetramerization,
e.g., Ir$^{3+}$/Ir$^{3+}$/Ir$^{4+}/$Ir$^{4+}$/... However, the experimentally
suggested structure exhibits a different pattern, W$^{5.5+x}/$W$^{5.5-x}%
/$W$^{5.5+x}/$W$^{5.5-x}/...,$ and experiment does not show any charge
disproportionation. That is to say, the structure proposed in Ref.
\cite{Hirai2013} still leaves uniform quasi-1D zigzag chains, running along
the crystallographic $b$ direction, which generate very metallic bands that
cannot open a gap even if the DFT bands are slightly off (see discussion below
and the corresponding band structure in Supplemental Materials (SM)). This
suggests that the real crystal structure for the low-T phase may have lower
symmetry than that proposed in Ref.~\cite{Hirai2013}. We will argue below that
the transition in question is actually a 3D Peierls transition, and that the
true LT structure encompasses four, not two cubic cells. 3D Peierls
transitions are extremely rare, but not impossible (for instance, the
nearest-neighbor $sp\sigma$ tight-binding model on the perovskite lattice at
some filling exhibits a perfect 3D nesting at $\mathbf{k=}[\frac{\pi}{a}%
,\frac{\pi}{a},\frac{\pi}{a}]$ \cite{Mattheiss1983}).

We will present below accurate DFT calculations of the electronic structure of
CsW$_{2}$O$_{6},$ and will show that upon including the spin-orbit interaction
(which appears essential for explaining the phase transition) it exhibits a
surprisingly simple Fermi surface (FS) with strong nesting for the three
equivalent wave vectors $\mathbf{Q}_{1}=[\frac{2\pi}{a},0,0],$ $\mathbf{Q}%
_{2}=[0,\frac{2\pi}{a},0],$ $\mathbf{Q}_{3}=[0,0,\frac{2\pi}{a}].$ This is
conducive to simultaneous condensation of the three corresponding charge
density waves (CDWs). Importantly, such condensation corresponds to a four-,
not eightfold supercell, as one may think. The phonon spectra indicates
instability exactly at these $\mathbf{Q}_{1},\mathbf{Q}_{2}$, and
$\mathbf{Q}_{3}$ points and optimizing crystal lattice starting from a
structure inspired by these phonon modes we obtained a lower symmetry
structure that opens a band gap. This structure is much lower in energy than
the published structure and yields a nonmagnetic ground state, which agrees
completely with experiment. \begin{figure}[t]
\includegraphics[width=0.8\columnwidth]{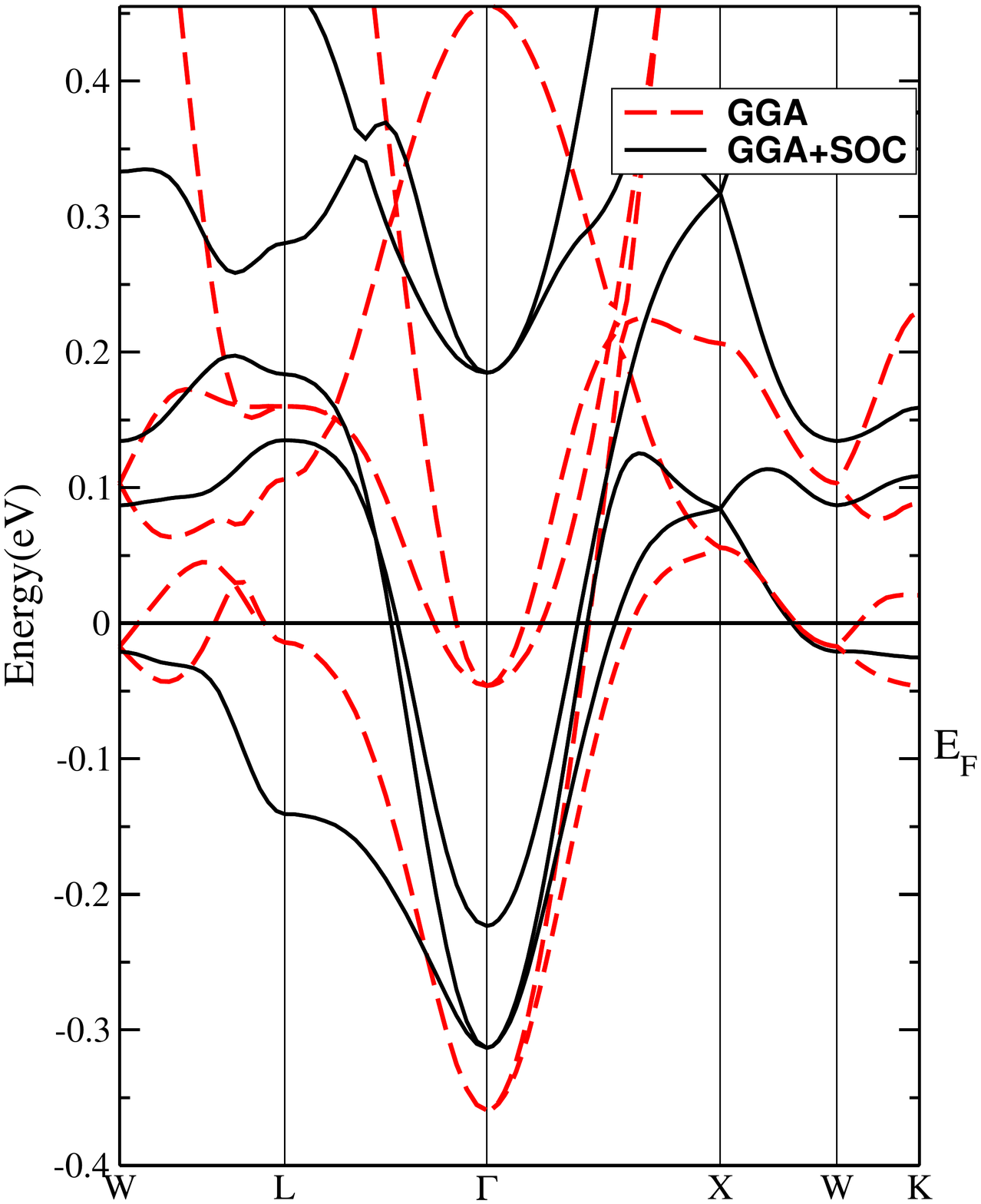}\caption{The band
structure obtained for the HT phase in the GGA and GGA+SOC calculations.
Positions of the high-symmetry points in the Brillouin zone are shown in Fig.
\ref{FS}.}%
\label{300}%
\end{figure}

Finally, we will discuss an intriguing implication of stabilizing the HT
structure at low temperature. We will argue that such a system could harbor a
highly unusual, and so far never observed, albeit theoretically predicted,
superconducting state.

{\textit{Computational results.} The DFT calculations were performed using the
full potential Linearized Augmented Plane Wave (LAPW) method (as implemented
in the Wien2k package\cite{Blaha2001}) with the Perdew-Burke-Ernzerhof (PBE)
exchange-correlation potential\cite{Perdew1996}. The noninteracting
susceptibility was computed on a fine mesh of 36000 $\mathbf{k}$-points in the
Brillouin zone. The optimization (atomic positions and cell shape were allowed
to change) of the low-T crystal structure was performed in the pseudopotential
VASP code\cite{Kresse1996} with the same type of the exchange-correlation
potential and taking into account SOC. Cs$-s$ and W$-p$ were treated as
valence states. We used cutoff 700 eV and the \textbf{k}-mesh 6$\times
$6$\times$6 for optimization. The maximal force in the converged structure was
less than 1 {meV/\AA }.}

In agreement with Ref.~\cite{Hirai2013}, we find that the GGA+SOC in the
proposed LT structure gives a strongly metallic ground state in a drastic
contrast to experiment. Moreover, this state turns out to be magnetic, with
small, but solid spin moments $m_{s}^{W1}\sim0.19\mu_{B}$ and $m_{s}^{W2}
\sim0.13\mu_{B}$. While these moments are further partially reduced by orbital
contributions $m_{o}^{W1}\approx m_{o}^{W2}\sim-0.05\mu_{B}$, they are still
non-negligible, which again stresses discrepancy with the experimental data.
Moreover, there are enormous atomic forces up to 0.8 eV/\AA , which make this
structure unstable in the GGA+SOC. As expected, including Hubbard correlations
within the GGA+U+SOC (we used $U=2$ eV and $J_{H}=0.5$ eV) only worsens the
situation, as moments begin to grow, while the system remains metallic.

In order to gain more insight into the physics of the LT phase we start by
analyzing the band structure of the HT cubic phase. Without SOC there are five
bands (Fig.~\ref{300}) crossing the Fermi energy ($E_{F}$). Two bands with
small dispersion cross $E_{F}$ in W-L, W-X and W-K directions, which results
in a large density of states (DOS) at $E_{F}$ and electronic instability.
There are also two bands crossing $E_{F}$ in the vicinity of the $\Gamma$
point. All these bands are mostly of the W $t_{2g}$ character.

The SOC dramatically modifies the band structure. The Fermi surface is
considerably simplified, and becomes canonically semi-metallic. The
unphysically large DOS is suppressed from $\sim$15 states/(eV f.u.) in
nonmagnetic GGA to $\sim$6 states/(eV f.u.) in GGA+SOC. Two bands are crossing
the Fermi level near $\Gamma,$ and one near X, forming, respectively, two
nearly degenerate electron pockets and three hole pockets per the reciprocal
cell. The former are nearly spherical, and the latter are more like rounded
parallelepipeds. \begin{figure}[t]
\includegraphics[angle=0,width=0.7\columnwidth]{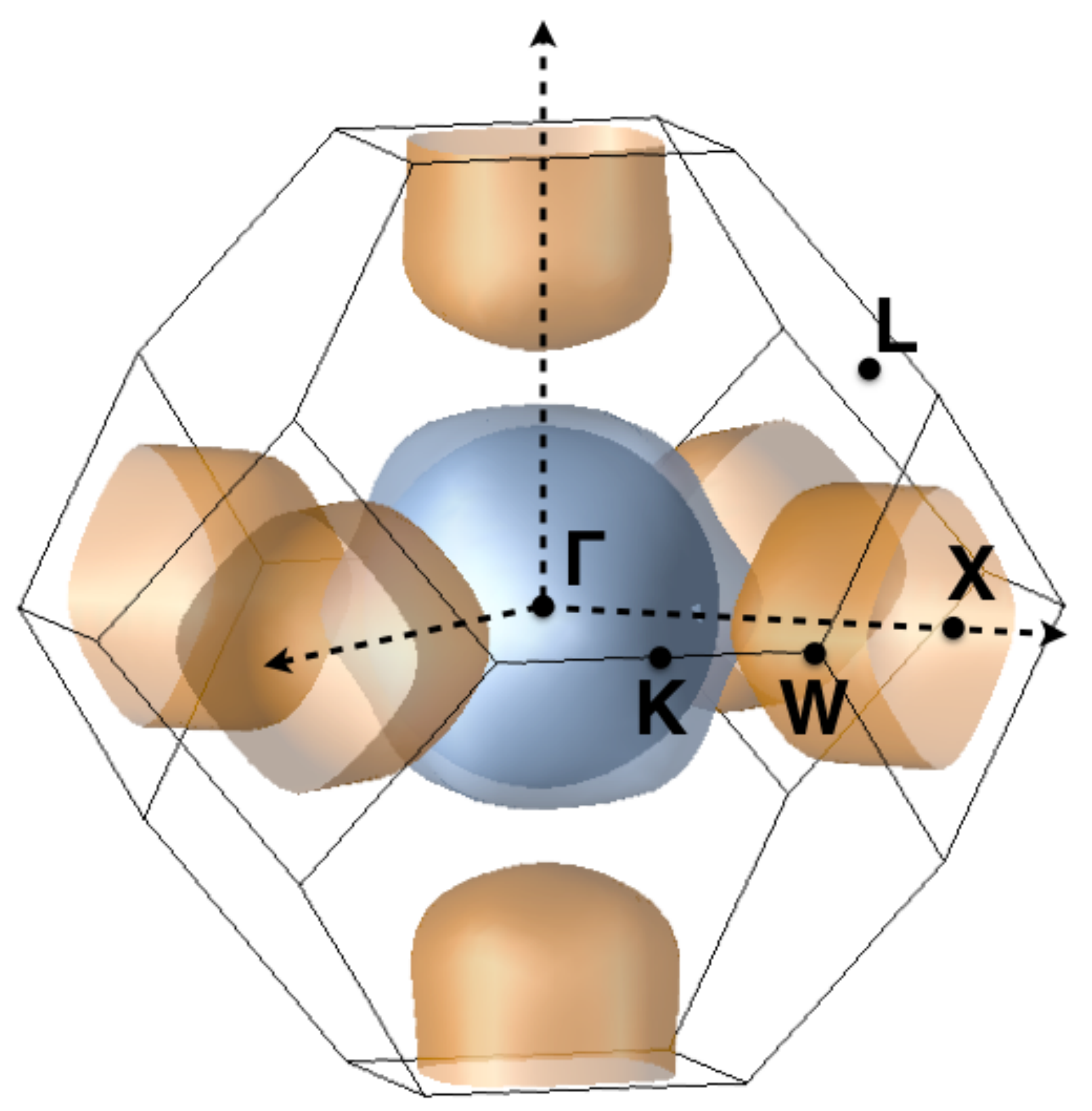}\caption{The Fermi
surface as obtained in the GGA+SOC calculation for the high-temperature phase.
}%
\label{FS}%
\end{figure}

This topology is prone to various instabilities. The energy mismatch when the
Fermi surfaces are shifted by the corresponding wave vector varies between 0
and $\sim$50 meV. Thus, energy can be gained by generating three simultaneous
CDWs that fold all three hole pockets right upon the electron pockets. As long
as the potential generated by the CDW is larger than $V\gtrsim50$ meV, a gap
of the order of $V$ will open, with a metal-insulator transition into a band
insulator phase
 \footnote{Note that this Fermi surface is topologically
consistent with an excitonic insulator instability\cite{Halperin1968}. Since
both CDW and $\mathbf{q} \neq0$ excitonic insulator break the same symmetries
and both can be diagrammatically described as instabilities in the
particle-hole channel \cite{Khomskii-book1}, we do not distinguish these in
the forthcoming discussion.}. 
Finally, as discussed in the next session,
arguably the most interesting instability this Fermi surface is conducive to
is an unconventional superconductivity.

Experimentally, it is clear that the second option is realized in the actual
material. Note that simultaneous condensation of the three CDWs in question
quadruples, but not octuples the unit cell. From the fact that the coordinates
of the X point of the Brillouin zone are $(2\pi/a,0,0)$ it is obvious that
going from the FCC Bravais lattice with the wave vectors $[a/2,a/2,0],$
$[a/2,0,a/2]$ and $[0,a/2,a/2]$ to the conventional cell with $[a,0,0],$
$[0,a,0]$ and $[0,0,a]$ corresponds to three CDWs with the X,Y and Z wave vectors.

\begin{figure}[t]
\includegraphics[width=0.95\columnwidth]{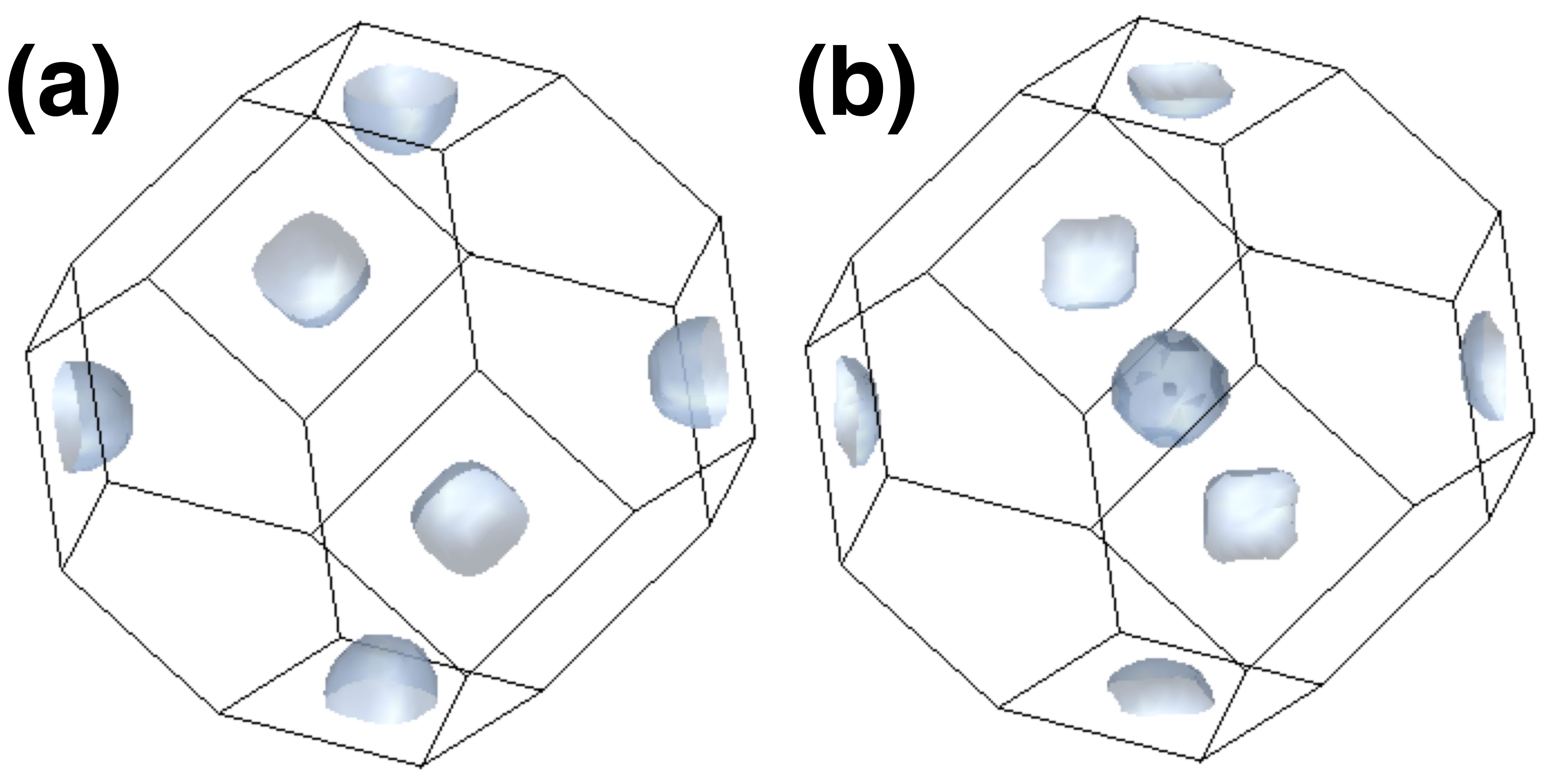}\caption{Real (a) and
imaginary (b) parts of the noninteracting susceptibility $\chi_{0}%
(\mathbf{q})$ calculated from the high-T band structure. Isosurfaces with
$\chi_{0}(\mathbf{q})= 92\%\chi_{0}^{max}$ are shown.}%
\label{chi}%
\end{figure}

Another way to look at this issue is to calculate, as it is often done,
\cite{Johannes2008a} the noninteracting susceptibility, neglecting the
\textbf{k}-dependence of the matrix elements, defined as
\begin{equation}
\chi_{0}(\mathbf{q})=\sum_{\mathbf{k},n,m}\frac{f_{\mathbf{k},n}%
-f_{(\mathbf{k}+\mathbf{q}),m}}{\varepsilon_{\mathbf{k},n}-\varepsilon
_{(\mathbf{k}+\mathbf{q}),m}}, \label{chi0}%
\end{equation}
where $f$ and $\varepsilon_{\mathbf{k},n}$ are the occupation numbers and
energies of the corresponding electronic states in the nonmagnetic GGA+SOC
calculation, $n$ and $m$ enumerate bands. The results are presented in
Fig.~\ref{chi} and clearly show peaks of $\chi_{0}(\mathbf{q})$ at the X,Y,
and Z points suggesting that the HT structure is unstable. While an account of
the interaction and momentum matrix elements may change the shape of full
susceptibility $\chi(\mathbf{q})$, it is unlikely to change the enhancement at
the ${X,Y,Z}$ wave vectors, since it is driven by the phase space factor
properly included already on the $\chi_{0}(\mathbf{q})$ level.

Keeping in mind these findings we performed calculation of the phonon spectra
using a relativistic linear response DFT technique \cite{Heid}
and found pronounced instabilities with largest negative frequencies exactly
at X point (see SM for details of the calculations and resulting phonon
spectrum). Condensing of these two phone modes yields structures with the
symmetry groups R32 and R\={3}m, respectively. An arbitrary linear combination
of these phonons gives the P4$_{1}$32 group. Optimizing the lattice in GGA+SOC
within any of these groups leads to lower (compared to the reported structure)
energies, but not insulating gaps. However, after checking possible subgroups
of the P4$_{1}$32 group we found that further lowering the symmetry to
P2$_{1}$2$_{1}$2$_{1}$ opens a gap, decreases the energy even further (by
$\sim$135 meV/f.u. compared to that proposed in Ref.~\cite{Hirai2013}) and is
nonmagnetic, which agrees with experimental data.

There are 16 W atoms in the unit cell in the optimized structure. Half of
these W form short W-W bonds (3.47\AA ), which results in an insulating ground
state with four W bands below $E_{F}$, occupied by all 8 available $5d$
electrons of 16 W$^{5.5+}$ ions (details of the crystal structure together
with corresponding band structure is given in SM). These short W-W bonds in no
case should be considered dimers (typical distance in W dimers are $\sim
$2.7\AA \cite{Johnstone2012}), but rather as a result of the CDW formation.
This is a consequence of the pyrochlore lattice, where oxygen is in between of
any two tungsten ions and prevents formation of real dimers. While average W-O
bond distances for four inequivalent W in the optimized structure are nearly
the same{ ($\sim$ 1.95 \AA )}, which agrees with results of
Ref.~\cite{Hirai2013}, there is a certain charge modulation on W sites (see
SM), consistent with formation of the CDW.

\textit{Superconductivity.} Materials close to a CDW instability often harbor
interesting superconductivity, which can either coexist with the CDW, or
emerge upon suppression of the latter. As discussed below, doped CsW$_{2}
$O$_{6}$ is a candidate for a highly unconventional superconducting state,
predicted by Agterberg et al in 1999\cite{Agterberg1999}, but never observed
(nor has any realistic candidate even been identified).

Let us briefly remind the reader of the essence of this work. Imagine the same
Fermi surface as depicted in Fig.~\ref{FS}, but without the electron pockets.
Let us further suppose that there is a pairing interaction that is stronger at
small momenta. When sufficiently strong Coulomb interaction is present in the
system (which is quite likely, given the small Fermi energy and thus small
logarithmic renormalization), the superconductivity is optimized if the phase
shift between the three inequivalent X-pockets is maximized, that is, equals
$2\pi i/3.$ Although not pointed out in the original paper\cite{Agterberg1999}%
, this state can be classified as \textquotedblleft$d+id$\textquotedblright,
being a combination of the two states with the $e_{g}$ symmetry,
$Y_{x^{2}-y^{2}}\pm iY_{3z^{2}-1},$ or ($Y_{2,2}+Y_{2,-2})/\sqrt{2}\pm
iY_{2,0}$, the same combination was discussed by van den Brink and Khomskii in
the context of the Jahn-Teller effect \cite{Khomskii2000a}.

One may argue that if an electron pocket is present, it has to have gap nodes.
This is true, but these nodes are only point nodes at the 8 directions
[$\pm1,\pm1,\pm1$], and thus the mean square gap ($\sqrt{155/512}\approx0.55)$
is only 15\% smaller than the maximal gap at ($\pm1,0,0),$ namely
$\sqrt{5/4\pi}.$ Thus, this highly unconventional state is quite viable and
may very well be realized in this material, if the high-temperature phase
could be stabilized. Given that, as shown above, the metal-insulator
transition in the real material is driven by a (spin-orbit induced) Peierls
instability, the most natural way to stabilize the HT phase is to dope it in
order to destroy nesting. Indeed a sister material, CsTaWO$_{6},$ occurs at
all temperatures in a cubic phase isostructural to the HT phase of CsW$_{2}%
$O$_{6}.$\cite{Knyazev200847}. This material is doped with 1/2 hole per 5$d$
metal ion, that is to say, W and Ta occur in the $d^{0}$ configuration, so
useless from the point of view of superconductivity. Synthesizing intermediate
materials of the composition CsW$_{1+x}$Ta$_{1-x}$O$_{6}$, $0<x<1,$ is more
promising and must be feasible. One caveat is in place: as any other $d-$wave
superconductivity, this $e_{g}+ie_{g}$ state would be sensitive to impurity
scattering, so the unavoidable W-Ta disorder may suppress or entirely destroy
this superconducting state. Other possibilities include (i) introducing Cs
vacancies, (ii) partially replacing O with N, or (iii) applying pressure (in
the latter case the nesting would likely be unchanged, but the elastic energy
penalty for the Peierls transition would increase).

\textit{Conclusions.} CsW$_{2}$O$_{6}$ appears to be a highly interesting
material, harboring quite unusual and even intriguing physics. So far it has
been barely studied experimentally and not investigated theoretically at all.
We hope that our work will stimulate further research.

Our main results are as follows: a close inspection of the experimentally
reported low-$T$ structure reveals that it cannot possibly open an insulating
gap, and at least a twice larger cell is needed, and that this structure
generates large forces in DFT, signaling that it is far from the lowest-energy
structure. Examining the high-$T$ structure, which had been unambiguously
established, we find that its fully-relativistic (spin-orbit coupling is
absolutely essential) Fermi surface exhibits an amazingly simple semi-metal
topology, with a good, well-visible in the calculated susceptibility,
electron-hole nesting. This nesting makes the high-$T$ structure unstable
against simultaneous formation of three charge density waves, running in the
three orthogonal crystallographic directions, i.e. susceptible to such a rare
phenomena as the 3D Peierls transition. Subsequent calculation of phonon
spectrum demonstrates that the largest negative phonon frequencies are exactly
at those points of the Brillouin zone, where the susceptibility diverges. The
crystal structure obtained by the lattice optimization using eigenvectors of
the lowest in energy phonon branches shows formation of short W-W bonds for
half of W atoms in the unit cell. CsW$_{2}$O$_{6}$ in this structure was found
to be insulating and nonmagnetic, fully consistent with available experimental
data. Having demonstrated that the observed transition is nesting-driven and
thus must be very sensitive to the band filling, we speculate that alloying
with Ta should suppress the CDW state rather rapidly.

Finally, we note that the calculated topology of the Fermi surfaces (in the HT
phase) is exactly the one required to realized an intriguing proposal of
Gor'kov and his collaborators about a 3D $d-$wave state, which breaks the time
reversal symmetry without being chiral, and can be characterized as a 3D
version of the famous ``$d+id$'' state (which in this case becomes
$e_{g}+ie_{g}).$

\textit{Acknowledgments.} We are grateful to R. Cava and D. Khomskii for
stimulating discussions and Z. Pchelkina for help with calculations. This work
was supported by Civil Research and Development Foundation via program
FSCX-14-61025-0, the Russian Foundation of Basic Research via Grant No.
16-32-60070, and the FASO (theme \textquotedblleft Electron\textquotedblright%
\ No. 01201463326 and act 11 contract 02.A03.21.0006). I.M. is supported by
ONR through the NRL basic research program.
The calculations were partially performed at Supercomputer cluster of the IMM UB RAS.

\bibliography{../library}

\end{document}